# PLC BASED UPGRADES FOR THE CAMD LINAC AND STORAGE RING CONTROL SYSTEM

P. Jines and B. Craft, Center for Advanced Microstructures and Devices, Louisiana State University, 6980 Jefferson Highway, Baton Rouge, LA 70806, USA

Abstract

Louisiana State University Center for Advanced Microstructures and Devices (CAMD) began a control system upgrade project in early 1997. At the time, the storage ring was controlled by a VAX/VMS system, primarily using CAMAC for I/O. The injector Linac was controlled by a separate VME/OS/9 system using custom I/O components. The storage ring control system has been replaced with a PC/Linux based system, utilizing the existing CAMAC subsystem [1]. In an effort to both migrate channels away from CAMAC, and integrate the Linac controls with the storage ring controls, a design for upgrades utilizing AutomationDirect.com PLCs was presented [2]. This paper will discuss the detailed design for the Linac PLC based control system, and discuss the status of PLC upgrades for the Vacuum system, Kicker controls, 2nd RF system, and utilities monitoring.

## 1 INTRODUCTION

When the CAMD control system upgrade project began in early 1997, CAMD was faced with the task of integrating two separate control systems, with no reliable communications method between the two. Although CAMD's LINAC and storage ring were both procured from Maxwell Laboratories via a single, fixed price contract, early in the design phase the Linac was subcontracted to CGR-MeV.

At the end of commissioning, the ring's control system was a VAX/VMS based system running Vista Controls VSystem software, utilizing CAMAC, GPIB, RS232, and Allen-Bradley Remote I/O communications. The Linac control system, however, was a VME/OS/9 system using custom I/O cards. While Maxwell and CAMD had specified TCP/IP access to the control system computer, subsystem acceptance tests had shown that this type of communications was not possible. If the system was "ping"ed via TCP/IP, the watchdog timers expired, and the software interlocks shut down the Linac.

The first control system to be upgraded was the storage ring control system. Initially, the VAX was replaced with a PC/Linux based system, preserving all existing hardware interfaces. Later phases of development were focused in two areas. The first was enhancing the controls capability, including adding additional control channels for a superconducting wiggler and additional corrector magnets, and increasing logging and operator automation. The second area was reducing the dependence on CAMAC and proprietary Allen-Bradley protocols, and allowing remote data acquisition. In this effort, GPIB and RS232/422/485 channels were given remote capability using National Instruments GPIB-ENET and Comtrol Rocketport Serial Hub Si remote ethernet based hubs. For all other upgrades, CAMD has standardized on AutomationDirect.com DL405 series PLCs.

## 2 PLC/CONTROL SYSTEM OVERVIEW

### 2.1 PLC Types

The AutomationDirect.com DL405 series PLCs have two different types of base controllers: the DL405 CPUs and the Ethernet Base Controller (EBC). The DL405 CPUs provide traditional CPU/Ladder Logic type control of the PLC, whereas the EBC eliminates the CPU, and provides a "pass thru" to allow the PC to control the PLC's I/O directly.

In combination with the DL405 CPU, the ethernet communications module allows the PC to read and write the PLC's memory without special communications software in the PLC. This led to the idea of using a "dual-port" memory model, where machine status and other parameters are communicated between the PLC and the control PC thru the PLC's memory. This provides a system with PLC type control, but PC based remote configurability.

### 2.2 Control System Integration

Channel information for CAMD's control system is stored in a PostgreSQL database, and is organized by device type, with device specific parameters as needed. As the CAMAC channel information is organized by crate, slot, channel address, and channel type, this seemed a logical approach to organize the PLC channel information. For the EBC type controller, this provided an exact match for the access method that the communications library provides. For the CPU based controllers, additional abstraction is necessary. The DL405 CPU architecture provides a flexible memory map: Inputs, outputs, control registers, and timers can be accessed directly, or as a "V memory" address. The ethernet communications protocol differentiates between different I/O and memory types by simply providing an address and a hex value for I/O or memory type. By combining this flexibility with the fact that the DL405 CPU's can provide a fixed starting I/O address for each slot (thereby eliminating I/O space size differences between a slot containing a 8 channel module and one containing a 64 channel module), a software mapping can be made between the "crate, slot, channel" view and the "V memory" address space. By adding fields to control the "base V memory" address, the PC code can be written without concern as to whether it is writing directly to I/O, or to a memory location, indicating that the PLC CPU will handle the I/O itself.

# 3 LINAC CONTROL SYSTEM UPGRADES

## 3.1 Current Linac Configuration and Migration Plan

The current Linac control system is based on a VME computer with custom I/O components. A custom scanner card in the VME crate using a dual-port memory map protocol communicates with the I/O cards via a 50-pin ribbon cable using a CGR-MeV in-house protocol. Each I/O card is homogenous in that all digital I/O is a TTL signal, and all analog I/O utilizes a 0-10V signal. Each card is given its address via external TTL voltages. The cards have a ribbon cable connector on one side, and connect to a 3U Eurocard backplane that handles all incoming and outgoing control signals. This backplane then connects to any signal specific hardware necessary, such as TTL to 24V contact relays, voltage to frequency converters, or directly to the devices controlled.

As all I/O is homogenous at the Eurocard backplane, this is the point at which the conversion from the old control system to the new PLC system will take place. Special cables have been fabricated to connnect D-shell DIN-rail breakout boxes to the 96-pin backplane connectors. Using this setup, the new control system can be tested, and the Linac can be reverted to the old control system simply by removing the adapter cables, and reinserting the I/O cards as before.

## 3.2 PLC Linac Control System Architecture

The Linac control system design utilizes simple, dedicated purpose PLCs. Analog interlocks and associated digital permits, control, and interlocks are all performed by separate PLCs or relays. This allows portions of the system to be distributed and enhanced without affecting the entire system.

Most digital Linac interlocks will be performed by dry relays. These will receive control inputs from EBC PLCs, and permits from smarter CPU based PLCs, which are responsible for all analog limit checking.

Temperatures, power supply readbacks, and other analog signals are monitored by dedicated "analog limit checking" PLCs. Each PLC is responsible for checking a group of analog inputs against upper and lower limits contained in its memory. The readbacks and limit values are all readable and/or writeable by the operator using the dual-port memory interface. In the event of an out of range indication, a digital output, or "permit", is turned off, which in turn is monitored by interlock PLCs or relays. The faults are latched, and must be reset by the operator before continuing.

Digital interlock PLCs and relays are used to monitor the analog "out of range" faults, and well as other incoming digital signals. These are responsible for removing permits in cases of low water flow, valve closures, power supply out of range conditions, or other indicators.

# 4 RING CONTROL SYSTEM UPGRADES

In addition to the Linac control system currently in progress, CAMD has also upgraded several ring systems to PLC based systems. These include Vacuum interlocks and monitoring, TSPs, and other digital control functions.

## 4.1 Vacuum Interlocks and Monitoring

CAMD has recently upgraded its vacuum system. There are significantly more Ion pump controllers and vacuum gauge controllers than in the older system. The decision was made to replace the old control system with a combination EBC/CPU PLC system. The system now consists of one identical rack per quadrant, each containing an EBC based PLC. These PLCs monitor the readbacks for each Ion pump controller and vacuum gauge controller in the quadrant. Contact closures from the vacuum gauge controllers are sent to a CPU based PLC, which contains the ladder logic and digital outputs necessary to close various ring valves in cases of high pressure conditions.

### 4.2 Titanium Sublimation Controllers

The initial TSP system for the CAMD storage ring contained only one sublimation controller, had no indication of the success or failure of the firing operation, and provided no indication as to how much current was applied to each filament. An upgrade was planned to provide one sublimation controller per quadrant, routing its output via a PLC controlled multiplexer, and to provide a graphical user interface with logging capability for the firing operation. For this activity, unused digital output channels from the vacuum system upgrade's EBC PLCs were used to control the multiplexer. The sublimation controllers' RS232 interfaces were controlled using the Comtrol Rocketport Serial Hub Si ethernet hubs. At the completion of the upgrade, the TSP system now can be run in one fourth of the time previously required, has a graphical indication of its activity, and has logging capability to monitor the vacuum system's performance.

### 4.3 Other PLC Upgrades

At present, CAMD is attempting to utilize CAMAC only for machine ramping functions, and to migrate all other controls to PLC based systems. In this effort, controls for both a second RF system being commissioned and the existing RF system are being converted from CAMAC to PLC control. Kicker controls have been converted from an Allen-Bradley PLC5 platform to the AutomationDirect.com platform. Monitoring for water plants, critical Linac temperatures and SF6 pressures are being provided by EBC based PLCs. Digital controls for new superconducting power supplies are being developed using the EBC platform. Finally, Ion Clearing Electrode controls have been converted to PLC based control.

## 5 CONCLUSION

CAMD has converted many existing controls to the AutomationDirect.com PLC systems. A new Linac control system is being developed using a combination of smart CPU based PLCs for interlock control, and less expensive EBC based control where only pure I/O is needed. At the conclusion of the upgrades, CAMD expects to have a consolidated, inexpensive, open and supportable system utilizing CAMAC where ramping is required, and non-proprietary PLCs for most other functions.

## REFERENCES

[1] B. Craft and P. Jines, 'The GNU Control System at CAMD,' Proceedings of the 1997 ICALEPCS, pp. 194-196, 1997
[2] P. Jines and B. Craft, 'The GNU Control System at CAMD: Part Two,' Proceedings of the 1999 ICALEPCS, pp. 118-120, 1999